\documentclass[12pt,a4paper]{article}
\usepackage{epsf}

\def\figuresize{15cm}

\begin{document}
\title{Localization and conductance in the quantum Coulomb glass
       \footnote{Dedicated to Prof.\ Michael\ Pollak on the
       occasion of his 75th birthday.}\\
       {\small Phil. Mag. {\bf B} 81, 1117 (2001)}
       }

\author{Thomas Vojta$^{1,2}$ and Michael Schreiber$^{1}$ \\ \small
 $^{1}$Institut f\"ur Physik, Technische Universit\"at Chemnitz,
 D-09107 Chemnitz, Germany \\
 \small Phone +49 371 531 3142, Fax +49 371 531 3143\\
 \small $^{2}$Theoretical Physics, University of Oxford, 1 Keble Road,
  Oxford OX1 3NP, UK}
\date{}
\maketitle

\addtolength{\baselineskip}{4pt}
\begin{abstract}
We consider the combined influence of disorder, electron-electron
interactions and quantum hopping on the properties of electronic systems
in a localized phase, approaching an insulator-metal transition.
The generic models in this regime are the quantum Coulomb glass and its
generalization to electrons with spin. After introducing these models
we explain our computational method, the Hartree-Fock based diagonalization.
We then discuss the conductance and compare spinless fermions and electrons.
It turns out that spin degrees of freedom do not play an essential role in the
systems considered. Finally, we analyze localization and decay of
single-particle excitations. We find that interactions generically
tend to localize these excitations which is a result of the Coulomb gap
in the single-particle density of states.
\end{abstract}

\section{Introduction}
The combined influence of quenched disorder and interactions in
electronic systems
remains one of the great unsolved problems in today's condensed matter
physics. Already disorder or interactions alone lead to a variety of
novel and intriguing physical phenomena like Anderson localization or
the Mott-Hubbard metal-insulator transition. Neither the disorder nor
the interaction problem can be considered completely understood today.

If both disorder and interactions are simultaneously present the behavior
becomes even more complex. A comparatively thorough understanding has been
achieved for systems with weak disorder and weak interactions in three
spatial dimensions which can be analyzed using conventional perturbative
methods (Lee and Ramakrishnan 1985).
The leading low-temperature behavior in these systems is not
influenced by the interactions: The electrons form a
dirty Fermi liquid (Belitz and Kirkpatrick 1999), a generalization of Landau's Fermi liquid concept to the disordered case.
At low temperatures, this implies a specific heat which is linear in the
temperature, a finite conductivity and a finite single-particle
density of states at the Fermi energy. However, the leading corrections
to the Fermi liquid theory already show nontrivial nonanalytic behavior due to
the presence of both disorder and interactions. These corrections which are
sometimes called Altshuler-Aronov singularities (Altshuler and Aronov 1985)
are, e.g., square-root
singularities in the density of states at the Fermi energy and in the
temperature or frequency dependence of the conductivity. This behavior has also
been observed experimentally in a variety of systems.
Based on these weak coupling results a quantum field theory was developed
and analyzed by means of renormalization group methods
(Finkelstein 1983, Belitz and Kirkpatrick 1994). This lead to a classification of the soft modes
in a disordered interacting electron system and the resulting identification
of different universality classes for the metal-insulator transition
(Belitz and Kirkpatrick 1994).

In two dimensions the situation is less clear even for weak disorder
and interactions. Perturbative methods analogous
to those used in three dimensions predict that non-interacting electrons
are always localized, and a metallic phase is impossible. In the presence
of interactions (and if additional symmetry-breaking terms are absent)
the renormalization group seems to predict an unconventional non-Fermi liquid
metallic state characterized by zero disorder but strong correlations
(Finkelstein 1983). However, this
prediction was based on an extrapolation of the renormalization group beyond its
region of validity. Therefore, it was widely believed that
a two-dimensional dirty electron system is generically insulating.
In recent years the two-dimensional case has attracted
renewed attention because experiments on Si-MOSFETs
(Kravchenko {\it et al.} 1994, Kravchenko {\it et al.} 1995) and other
systems with very low electron density show signatures of a metal-insulator
transition, in contradiction to the orthodox theory. Up to now
this transition, if any, is not understood.
In a number of theories new quantum states of matter have been
postulated, either non-Fermi liquid metals
(Castellani {\it et al.} 1998, Si and Varma 1998, Benenti {\it et al.} 1999,
Chakravarty {\it et al.} 1999, Denteneer {\it et al.} 1999) or unusual
superconductors (Zhang and Rice 1997, Belitz and Kirkpatrick 1998,
Philips {\it et al.} 1998). However, other explanations
like temperature-dependent disorder (Altshuler and Maslov 1999)
or impurity screening (Gold and Dolgopolov 1986, Das Sarma and Hwang 1999,
Klapwijk and Das Sarma 1999)
are more in line with the orthodox view. They imply that the seeming MIT is a transient
phenomenon, and the true ground state is always an insulator.

Most of the theoretical work discussed above is concerned with the metallic
phase and with
approaching the metal-insulator transition from the metallic side. To a large
extent
independently, the behavior of interacting disordered electrons in the
localized regime has been investigated for more than 30 years.
Pollak (1970) suggested that the Coulomb interaction leads to a reduction
of the single-particle density of states at the Fermi energy. Efros and
Shklovskii (1975) derived an approximate expression for the density
of states in this soft gap which they called the Coulomb gap. Later, a large amount
of research went into the question, how hopping transport in the localized
regime is influenced by the Coulomb interaction (see, e.g. the review
articles in Efros and Pollak (1985)). Despite these efforts,
the problem cannot be considered completely solved.

The interest in the interplay between Anderson localization and interactions
resurged with the work of Shepelyansky (1994)
on the localization of just two interacting particles. He suggested that
even for a repulsive interaction the two particles can form a pair whose
localization length is much larger than that of a single particle.
By now it is established, that such an enhancement of the localization length
indeed exists, even though the precise dependence of the pair localization
length on disorder and interaction strength has not been conclusively
determined yet (R\"omer {\it et al.} 2001).
However, the experimentally more significant case is a finite
electron density in the thermodynamic limit as opposed to just two particles.
Since a systematic analytic approach to the insulating phase of a
disordered electron system does not exist, most of the work on this problem
has relied on numerical methods, e.g., the density-matrix renormalization
group in one-dimensional systems (Schmitteckert {\it et al.} 1998) and exact or semi-exact diagonalization
methods in higher dimensions (Berkovits and Avishai 1996, Berkovits {\it et al.}
2001).

In this paper we investigate the transport and localization properties of
disordered interacting electrons in the localized regime using the Hartree-Fock
based diagonalization method (Vojta {\it et al.} 1999),
an effective numerical method for disordered
many-particle systems. The paper is organized as follows.
In section \ref{sec:CCG} we introduce the generic model for the
description of disordered interacting electrons in the extremely localized
limit, the classical Coulomb glass, and discuss its basic properties.
In section \ref{sec:QCG} we introduce quantum hopping and spin degrees
of freedom into the model.
The resulting quantum Coulomb glass and its spin generalization
are the prototypical models for
disordered electrons in the insulating phase. Section \ref{sec:HFD}
is devoted to explaining our numerical method, the Hartree-Fock based
diagonalization. We present and discuss our results on transport
and localization in section \ref{sec:RES}.

\section{Disorder and interactions in the localized limit}
\label{sec:CCG}

In the extremely localized or classical limit the kinetic energy of the electrons
can be neglected compared to the potential energy and the electron-electron
interaction. In a doped semiconductor this corresponds, e.g., to a
vanishing concentration of donors or acceptors. In this limit
the electrons behave like classical point charges, and the spin degrees
of freedom do not play any role. The generic model
for this regime is the classical Coulomb glass model (Efros and Shklovskii 1975)
which consists of
classical point charges in a random potential which interact via Coulomb interaction.
The model is defined on a regular hypercubic lattice with one state on each of its
$\mathcal{N}=L^d$ ($d$ is the spatial dimensionality) sites.
The system is occupied by $N= K \mathcal{N}$ spinless
fermions ($0\!<\!K\!<\!1$). To ensure charge neutrality
each lattice site carries a compensating positive charge of  $Ke$. The Hamiltonian
of this classical Coulomb glass reads
\begin{equation}
H_{\rm cl} = \sum_i (\varphi_i - \mu) n_i + \frac{1}{2}\sum_{i\not=j}(n_i-K)(n_j-K)U_{ij}~.
\end{equation}
Here $n_i$ is the occupation number of site $i$ and $\mu$
is the chemical potential. The Coulomb interaction $U_{ij} = e^2/r_{ij}$
remains long-ranged since screening breaks down in the insulating phase.
We parametrize the interaction by its nearest-neighbor value $U$.
The boundary conditions are periodic and the interaction is treated in
the minimum image convention (which implies a cut-off at a distance of $L/2$).
The random potential values $\varphi_i$ are chosen
independently from a box distribution of width $2 W_0$ and zero mean.
In the following we set the disorder strength $W_0$ to 1 which fixes the
energy scale.

One of the most important properties of the classical Coulomb glass is
the reduction of the single-particle density of states at the Fermi energy
$\epsilon_F$ (Pollak 1970). This soft Coulomb gap is of the form
(Efros and Shklovskii 1975)
\begin{equation}
 g(\epsilon) \sim |\epsilon - \epsilon_F|^\alpha~.
\end{equation}
Using a self-consistent equation of mean-field type
the exponent was determined to be $\alpha$ to be 1 in two dimensions
and 2 in three
dimensions. Numerical simulations usually give slightly larger values
(M\"obius {\it et al.} 1992), the latest being
1.7 for 2D and 2.7 for 3D (Sarvestani {\it et al.} 1995).

Many-particle excitations like particle-hole excitations have weaker gaps
or no gap at all.\footnote{We note that Efros and Shklovskii (1975) considered
the energy required for a {\it particle-hole} excitation to
derive the Coulomb gap in the {\it single-particle} density of states.}
The influence of these gaps on the transport properties
of the classical Coulomb glass has been the subject of a long controversy.
Since the electrons are localized, transport must occur via hopping.
Efros and Shklovskii generalized Mott's variable-range hopping theory
(Mott 1968)
to cases with a non-constant single-particle density of states. According
to this theory the conductivity varies like $\sigma \sim \exp[-(T_0/T)^{1/2}]$.
While this behavior is indeed observed in many relevant systems, the
quantitative values for hopping distance and energy often do not work out.
The above theory was criticised by Pollak
(for reviews see Efros and Pollak 1985, Pollak 1992)
for essentially being a
single-particle transport theory.  Such a theory should not be adequate in an
interacting system since the excitations responsible for transport are
particle-hole or more complex excitations which do not have such a strong Coulomb gap.
So far no completely satisfactory theory has been developed.
We will show in section \ref{sec:RES} that similar phenomena also exist for quantum
transport: Single-particle and particle-hole excitations can have very
different localization properties.

\section{From the classical to the quantum Coulomb glass}
\label{sec:QCG}

If the electrons are not completely localized, each at one impurity site,
anymore, the overlap between states at different sites cannot be neglected
even if the system is still in the insulating phase.
To describe this regime, we add a kinetic energy, i.e. quantum
hopping matrix elements of
strength $t$ between nearest neighbors, to the Hamiltonian:
\begin{equation}
H_{\rm qu} =  -t  \sum_{\langle ij\rangle} (c_i^\dagger c_j + c_j^\dagger c_i) + H_{\rm cl},
\label{eq:QCG}
\end{equation}
where $c_i^\dagger$ and $c_i$ are the fermion creation and annihilation operators
at site $i$, respectively,  and the sum runs over all pairs of nearest neighbor sites.
Periodic boundary conditions are also used for the hopping part of the Hamiltonian.
In the limit $t \rightarrow 0$ the model (\ref{eq:QCG}) reduces to the classical
Coulomb glass, for vanishing Coulomb interaction but finite overlap it reduces to the
usual Anderson model of localization.
The model defined in (\ref{eq:QCG}) is often called the quantum Coulomb glass
(Efros and Pikus 1995, Talamantes {\it et al.} 1996, Epperlein {\it et al.}
1997).

However, in the presence of quantum hopping the spin-degrees of freedom
of the electrons cannot simply be neglected, and using the quantum Coulomb
glass Hamiltonian (\ref{eq:QCG}) involves an uncontrolled
approximation. Even in the insulating phase, a complete description of
disordered interacting electrons therefore requires a generalization of
the quantum Coulomb glass model to electrons with spin.
We again consider a hypercubic lattice with $\mathcal{N}=L^d$ sites,
now with two states (spin up and down) per site.
The system contains $N=N_\uparrow +N_\downarrow=2K \mathcal{N}$
electrons ($0\!<\!K\!<\!1$) and has a compensating positive charge of
$2Ke$ on each lattice site. The Hamiltonian is given by
\begin{eqnarray}
H &=&  -t  \sum_{\langle ij\rangle, \sigma} (c_{i\sigma}^\dagger c_{j\sigma}
      + c_{j\sigma}^\dagger c_{i\sigma})+
       \sum_{i,\sigma} \varphi_i  n_{i\sigma} \\
&&+~\frac{1}{2}\sum_{i\not=j,
       \sigma,\sigma'}U_{ij}~(n_{i\sigma}-K)(n_{j\sigma'}-K)\nonumber\\
&&+~ U_{\rm H} \sum_{i} n_{i\uparrow} n_{i\downarrow}\nonumber
\label{eq:Spin_QCG}
\end{eqnarray}
where $c_{i\sigma}^\dagger$, $c_{i\sigma}$, and $n_{i\sigma}$ are the creation,
annihilation and occupation number operators for electrons at site $i$
with spin $\sigma$.
Two electrons on the same site interact via a Hubbard interaction $U_{\rm H}$.

\section{Hartree-Fock approximation and Hartree-Fock \\based diagonalization}
\label{sec:HFD}

The simulation of a disordered quantum many-particle system like the quantum
Coulomb glass or its spin generalization poses formidable problems for any numerical
approach. On the one hand, the dimension of the many-particle Hilbert space
grows exponentially with the system size, restricting the simulation to very
small sizes even for clean many-particle systems.
On the other hand, the presence of disorder requires the simulation of
many samples with different disorder configurations in order to
obtain averages or typical values of physical quantities. The problem is
even more severe when non-self-averaging quantities like the conductance
are considered. Here one should in principle study the entire distribution
function of an observable instead of just one characteristic value.
In the case of the quantum Coulomb glass, an additional complication is
produced by the long-range character of the Coulomb interaction
which has to be retained, at least for a correct description of
the insulating phase.

In the literature, several different methods have been used to study
the quantum Coulomb glass and related problems. However, most of them
are hampered by severe problems.
Exact diagonalization (Dagotto 1994, Epperlein {\it et al.} 1998)
works only for very small systems
(with up to about $4 \times 4$ lattice sites for electrons with spin).
For one-dimensional systems the density-matrix renormalization
group method (White 1998) is a very efficient tool to obtain the low-energy properties.
It is, however, less effective in higher dimensions; and it is also not capable of handling
the long-range Coulomb interaction which is important in the insulating phase.
Quantum Monte-Carlo methods (von der Linden 1992) are another means of simulating disordered many-particle
systems. They are very effective for bosons at finite temperatures. Very low temperatures are,
however, hard to reach. Moreover, simulations of fermions suffer from the notorious sign problem
(although this turned out to be less severe in the presence of disorder).

In this paper we use two different numerical methods to investigate
the quantum Coulomb glass. For the calculation of static properties
like the single-particle density of states we use a disordered-Hartree-Fock
approximation (Epperlein {\it et al.} 1997). We decouple the
Coulomb and Hubbard interaction terms in the Hamiltonian following the
recipe $c_{i\sigma}^\dagger c_{j\sigma'}^\dagger c_{j\sigma'} c_{i\sigma}
\to c_{i\sigma}^\dagger c_{i\sigma} \langle c_{j\sigma'}^\dagger c_{j\sigma'}\rangle
-  c_{i\sigma}^\dagger c_{j\sigma'} \langle c_{j\sigma'}^\dagger c_{i\sigma}\rangle$.
The resulting disordered single-particle Hamiltonian is diagonalized
numerically in a self-consistency cycle, until convergence is reached for the
expectation values $\langle \ldots \rangle$ with respect to the ground state.
This calculation results in an orthonormal set of single-particle
Hartree-Fock states.
Although decoupling the interaction terms involves an uncontrolled approximation
the method gives reasonable results for the density of states, as we have shown
by comparison with exact diagonalization (Epperlein {\it et al.} 1998).
Its main advantage is that comparatively
large systems of several thousand sites can be handled, because effectively only
a single-particle problem has to be solved.

For quantities defined by time correlation functions
like the conductance the Hartree-Fock approximation gives
very poor results  (Epperlein {\it et al.} 1998).\footnote{In principle one
should use the time-dependent
Hartree-Fock method which would certainly
improve the results but is considerably more expensive numerically.}
In recent years we have therefore developed and used an alternative method.
It improves the Hartree-Fock approximation by
diagonalizing the many-particle Hamiltonian in the subspace of the
Hilbert space spanned by
the low-energy Slater states calculated in the Hartree-Fock approximation.
This method,
the Hartree-Fock based diagonalization (HFD),
is analogous to the quantum-chemical configuration interaction (see, e.g.,
Fulde 1995)
approach but adapted for disordered lattice models.

The HFD method consists of 3 steps, which have to be carried out separately
for each disorder configuration considered:
(i) Solve the Hartree-Fock approximation of the Hamiltonian, as discussed above,
    which gives a set of single-particle Hartree-Fock states.
(ii) Construct many-particle Slater determinants from these states, and
     use a Monte-Carlo algorithm based on thermal cycling (M\"obius {\it et
     al.} 1997)
     to find a certain number $B$ of the lowest-in-energy Slater states.
(iii) Transform the Hamiltonian and the observables to the basis formed by these
     Slater states and diagonalize the Hamiltonian in the Hilbert space
     they span. Calculate the  observables.

The efficiency of the HFD method is due to the fact that the Hartree-Fock
states are comparatively close in character to the exact eigenstates in the entire
parameter space, as we have shown in comprehensive test runs.
Thus the method works well for all parameters while related methods
based on non-interacting or classical eigenstates
(Efros and Pikus 1995, Talamantes {\it et al.} 1996)
instead of Hartree-Fock states are restricted to small parameter regions.
We have carried out extensive tests of the method and investigated how the
quality of the results depends on the dimension of the restricted Hilbert space
(i.e., the number of Slater states) used in step (iii) (Vojta {\it et al.} 1999).
For a relatively small system of 16 sites and 8 spinless fermions we compared
the HFD to exact diagonalization results. We found that already
a basis size of $B=100$ Slater states (the total dimension of the Hilbert space
is 12870 in this case) yields the ground state energy with an
relative error less than $10^{-5}$
and the respective occupation numbers with an error less than $10^{-2}$.
For larger systems we checked the convergence by systematically increasing $B$
until the results did not change within the desired accuracy. For systems
with up to 64 sites we estimated that a few hundred to 2000
Slater states are sufficient for the questions studied here.

\section{Physical properties of the quantum Coulomb glass}
\label{sec:RES}

In our previous work we have already investigated the transport properties of
spinless fermions in one (Schreiber {\it et al.} 1999),
two (Vojta {\it et al.} 1998b) and three (Vojta and Epperlein 1998)
dimensions in some detail.
Here we concentrate on the generalized quantum Coulomb glass with spin
degrees of freedom in two dimensions. The presence of the spin degrees
of freedom strongly increases the dimension of the Hilbert space. Therefore,
for most production runs we have studied systems of $4\times 4$ sites.
The numerical effort is equivalent to that for a system of 32 sites
in the case of spinless fermions.

We first investigate the conductance which is the most accessible observable
in a real transport experiment. Theoretically, it can be
obtained from linear-response theory. It is essentially determined by the
current-current correlation function of the ground state.
The real (dissipative) part of the conductance (in units of $e^2/h$)
at frequency $\omega$
is given by the Kubo-Greenwood formula (Kubo 1957, Greenwood 1958),
\begin{equation}
 {\rm Re} ~ G^{xx}(\omega) = \frac {2 \pi^2}  {\omega} \sum_{\nu} |\langle 0 | j^x|\nu \rangle |^2
     \delta(\omega+E_0-E_{\nu}).
\label{eq:kubo}
\end{equation}
$j^x$ is the $x$ component of the current operator and $|\nu\rangle$ is an eigenstate
of the Hamiltonian. Equation (\ref{eq:kubo}) describes an isolated system while
 in a realistic d.c.\ transport experiment
the sample is connected to contacts and leads. This results in a finite life time $\tau$
of the eigenstates leading to an inhomogeneous broadening $\gamma = \tau^{-1}$
of the $\delta$ functions in (\ref{eq:kubo}) (Datta 1997). To suppress
the discreteness of the spectrum of a finite system, $\gamma$ should be
at least of the order of the single-particle level spacing.
In our still comparatively small systems this requires
large $\gamma \ge 0.05$. We tested different $\gamma$
and found that the conductance {\em values} depend on $\gamma$ but the
qualitative results do not.

\begin{figure}
  \epsfxsize\figuresize
  \centerline{\epsffile{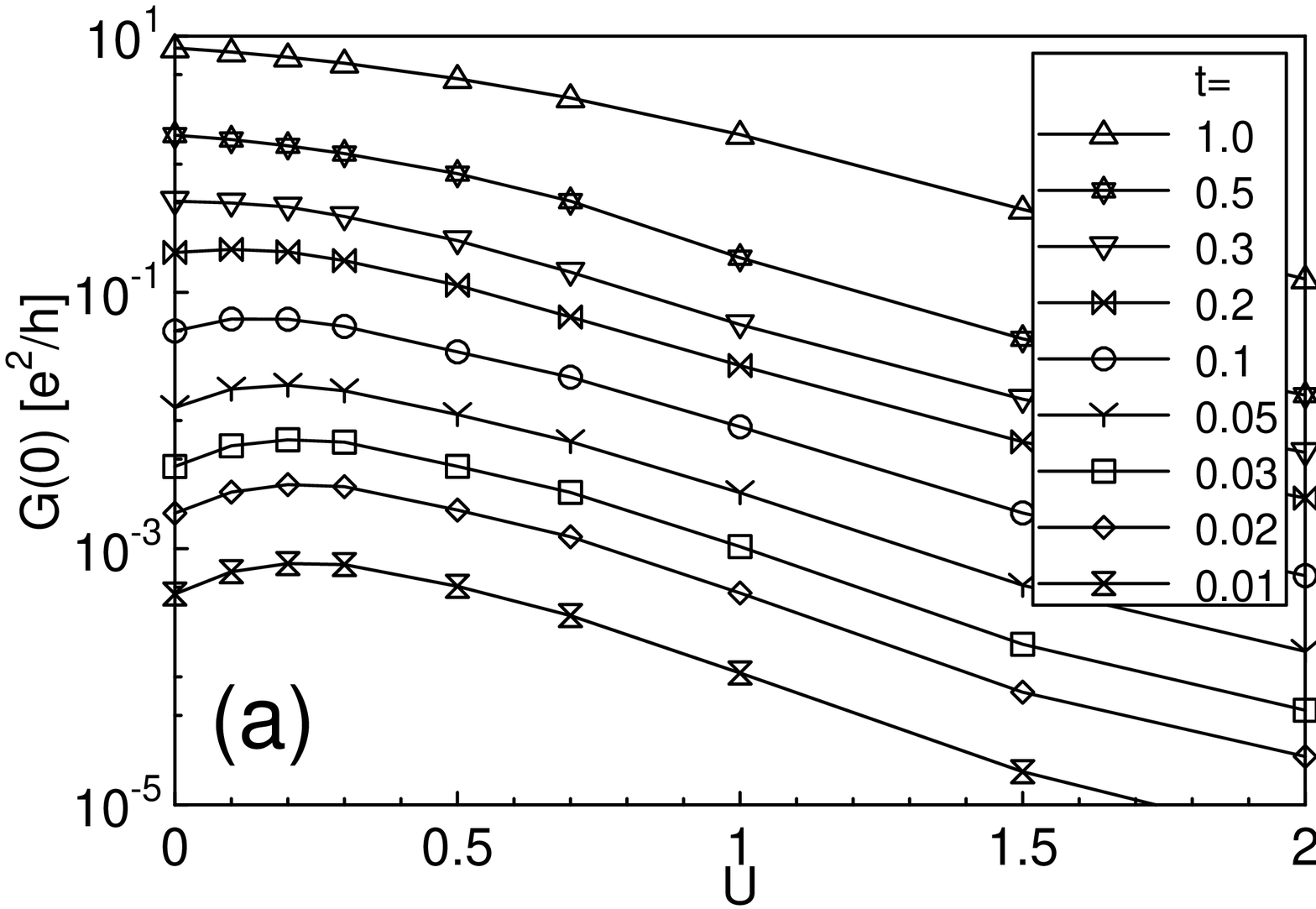}}
  \epsfxsize\figuresize
  \centerline{\epsffile{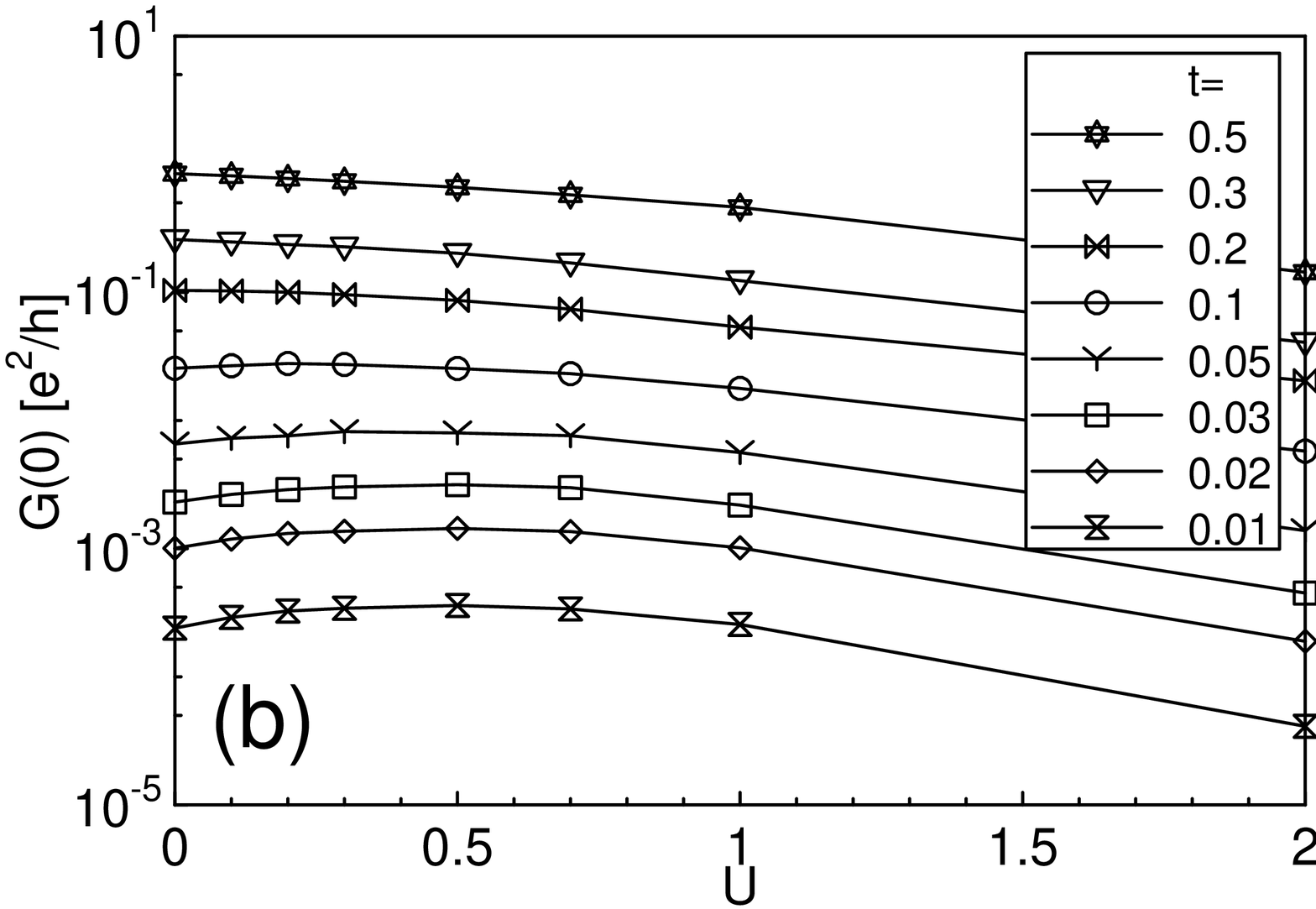}}
  \caption{d.c. conductance $G(0)$ as a function of interaction $U$
      for a system of $4 \times 4$ sites occupied by
      (a) 8 spin-up and 8 spin-down electrons and (b) 8 spinless fermions.
      The Hubbard energy is $U_H=U$, and the HFD basis size $B=500$.
      The inhomogeneous broadening
      is $\gamma=0.0625$.}
  \label{Fig:conductance}
\end{figure}

Figure \ref{Fig:conductance} shows the typical conductance values
of a system of $4 \times 4$ lattice sites at half filling as
a function of the interaction $U$ for different hopping matrix elements $t$.
Panel (a) shows results for an occupation of 8 spin-up and 8 spin-down
electrons while panel (b) is for 8 spinless fermions.
Since the logarithm of the conductance rather than the
conductance itself is a self-averaging quantity in a disordered system,  we calculate
the typical conductance by averaging the logarithms of the conductances of 1000
(400 in the spinless case) disorder configurations.
Both graphs show the same qualitative behavior:
For large kinetic energy $t$ the interactions always reduce the
d.c. conductance, while for small $t$, i.e. in the localized regime,
small and moderate interactions significantly enhance the d.c. conductance.
For larger interaction strength the conductance drops,
indicating the crossover to a Wigner crystal or Wigner glass.

\begin{figure}
  \epsfxsize\figuresize
  \centerline{\epsffile{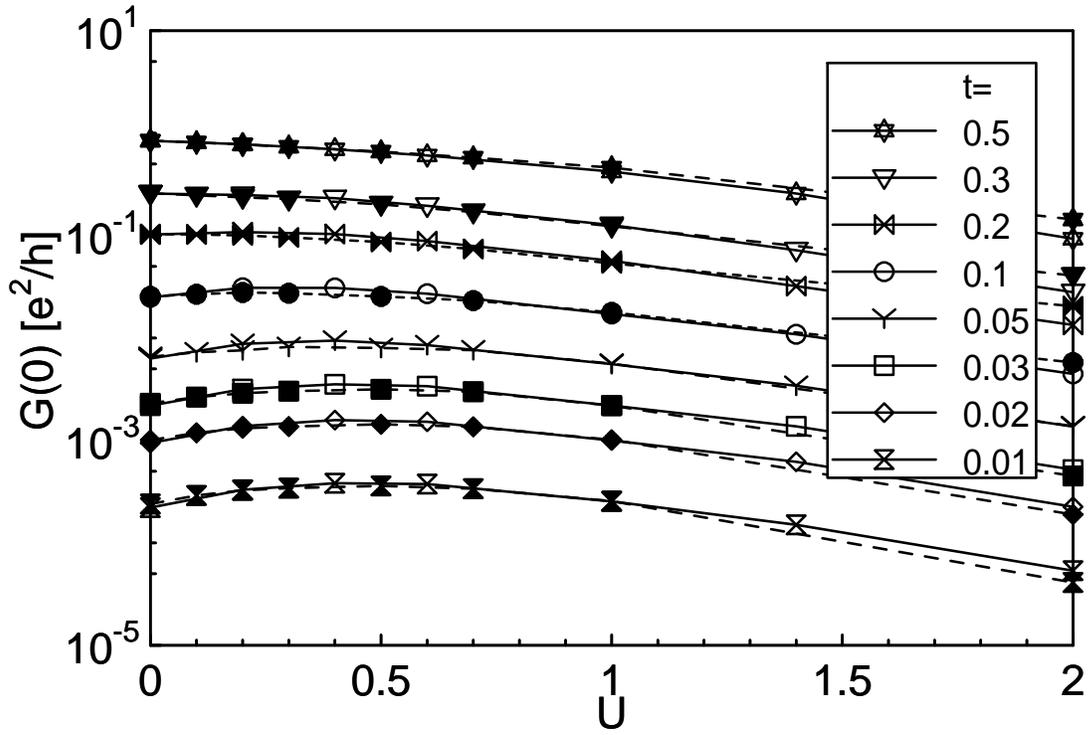}}
  \caption{Comparison of the d.c. conductance of spinless fermions
  (full symbols) and electrons with spin (open symbols). In the spin
  case the conductance has been scaled by 1/2 and the interaction
  strength by 2. The parameters are as in Fig. \protect\ref{Fig:conductance}.}
  \label{Fig:compare}
\end{figure}

A closer comparison of the cases with and without spin reveals a number
of interesting similarities and differences.
Without interactions ($U=0$) the conductance of the spinless fermions
is exactly half of that of the electrons with spin. This just reflects the
fact that at zero interactions spin-up and spin-down electrons decouple.
The contributions of these two subsystems to the conductance are identical
and identical to that of spinless fermions. With increasing interaction
the conductance increases faster for electrons than for spinless
fermions, but it also falls off faster. One possible reason is that in the
case with spin there is twice as much charge in the system as in the spinless case,
and thus effectively the Coulomb interaction is twice as strong.
To explore this we directly compare the two results
in figure \ref{Fig:compare}
after rescaling the conductance of the electron system by 1/2
and the interaction strength by 2. The two sets of curves nicely fall on top
of each other within the statistical accuracy.
Therefore we conclude that for the systems considered
the Coulomb interaction
plays the essential role for the delocalizing tendency for weak
interactions as well as for the localizing tendency for strong interactions.
The spin degrees of freedom do not seem to be important.

\begin{figure}
  \epsfxsize\figuresize
  \centerline{\epsffile{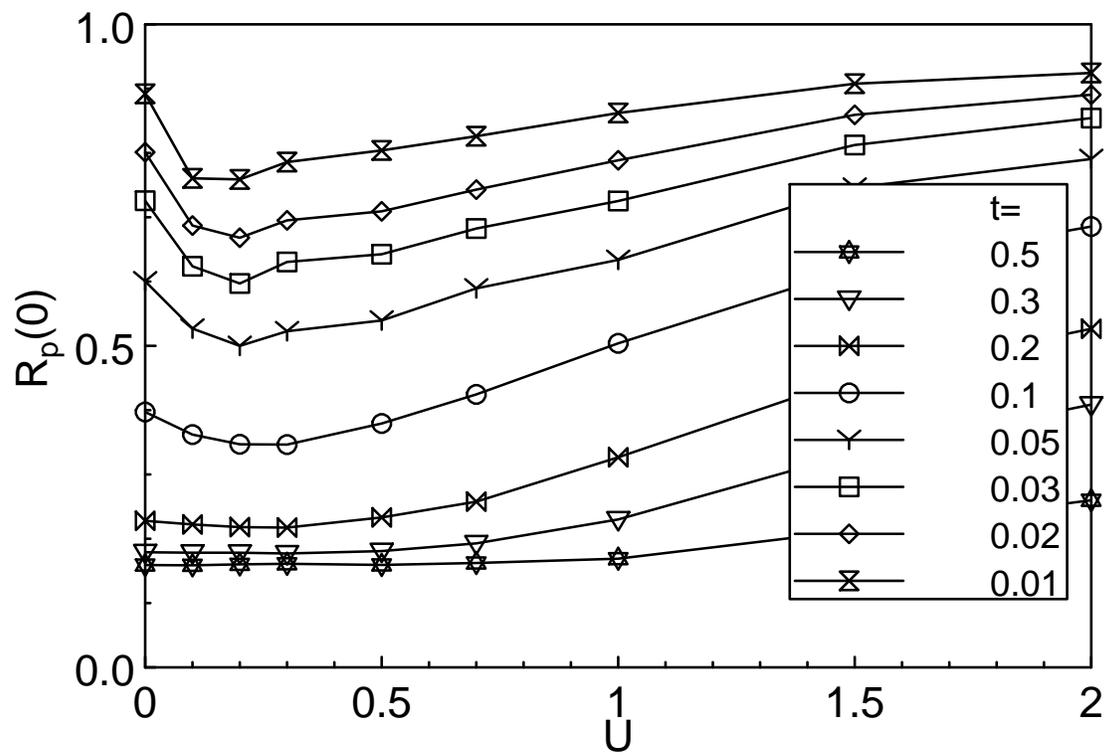}}
  \caption{Single-particle return probability at the Fermi energy, $R_\sigma(0)$,
  for a system of $4 \times 4$ sites occupied by
  8 spin-up and 8 spin-down electrons; the other parameters are
  as in Fig. \protect\ref{Fig:conductance}.}
  \label{Fig:return}
\end{figure}

We now turn to the single-particle localization properties. In the case
of non-interacting electrons they can be characterized by the inverse
participation number of the single-particle states with respect to the
site basis states. This quantity is identical to the return probability
of the single-particle excitations (Economu and Cohen 1970),
\begin{equation}
R_\sigma(\omega) =  \frac 1 {g(\epsilon)} \frac 1 {\mathcal{N}} \sum_{i} \lim_{\delta \to 0} \frac \delta \pi \, G_{i\sigma i\sigma}^R(\omega + i \delta)
     \, G_{i\sigma i\sigma}^A(\omega - i \delta).
\label{eq:return}
\end{equation}
Here $G_{i\sigma j\sigma'}^{R,A}(\omega)$ are the retarded and advanced
single-particle Greens functions.
In contrast to the single-particle participation number the return probability
is also well defined for interacting systems (Vojta {\it et al.} 1998a).
In figure \ref{Fig:return} we show the return probability as a function of the
interaction strength for different hopping matrix elements $t$.
Its behavior is very similar to that of the conductance.
For large kinetic energy $t$ the interactions always increase the
return probability, i.e. they lead to stronger localization. For small $t$,
small and moderate interactions significantly decrease the return probability
and thus lead to weaker localization.

\begin{figure}
  \epsfxsize\figuresize
  \centerline{\epsffile{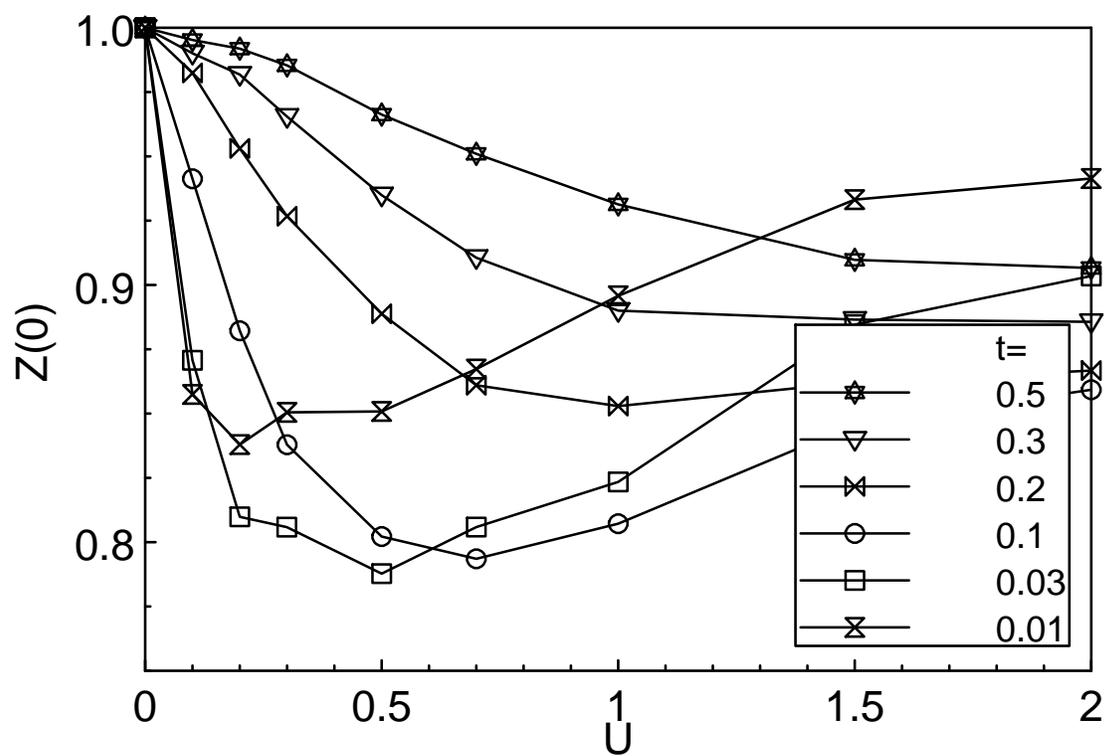}}
  \caption{Single-particle survival probability at the Fermi energy, $Z_\sigma(0)$;
  parameters as in Fig. \protect\ref{Fig:return}.}
  \label{Fig:quasi}
\end{figure}

However, in an interacting system the return probability (\ref{eq:return})
entangles localization and decay information, since $R_\sigma$ is decreased
from 1 not only by delocalization but also by decay of the
quasi-particles. In order to extract the decay of the quasi-particles
we consider the survival probability
\begin{equation}
Z_\sigma(\omega) = \frac 1 {g(\epsilon)} \frac 1 {\mathcal{N}} \sum_{ij} \lim_{\delta \to 0} \frac \delta \pi \, G_{i\sigma j\sigma}^R(\omega + i \delta)
     \, G_{j\sigma i\sigma}^A(\omega - i \delta),
\label{eq:survive}
\end{equation}
which sums the transmission probabilities of a single-particle excitation
to all sites. The survival probability at the Fermi energy is related to
the square of the quasiparticle weight.
Our results for the survival probability are shown in Fig. \ref{Fig:quasi}.
For non-interacting electrons $Z_\sigma=1$ by definition since single-particle
excitations do not decay. Weak interactions lead to a reduced survival
probability but for stronger interactions it approaches 1 again.

\begin{figure}
  \epsfxsize\figuresize
  \centerline{\epsffile{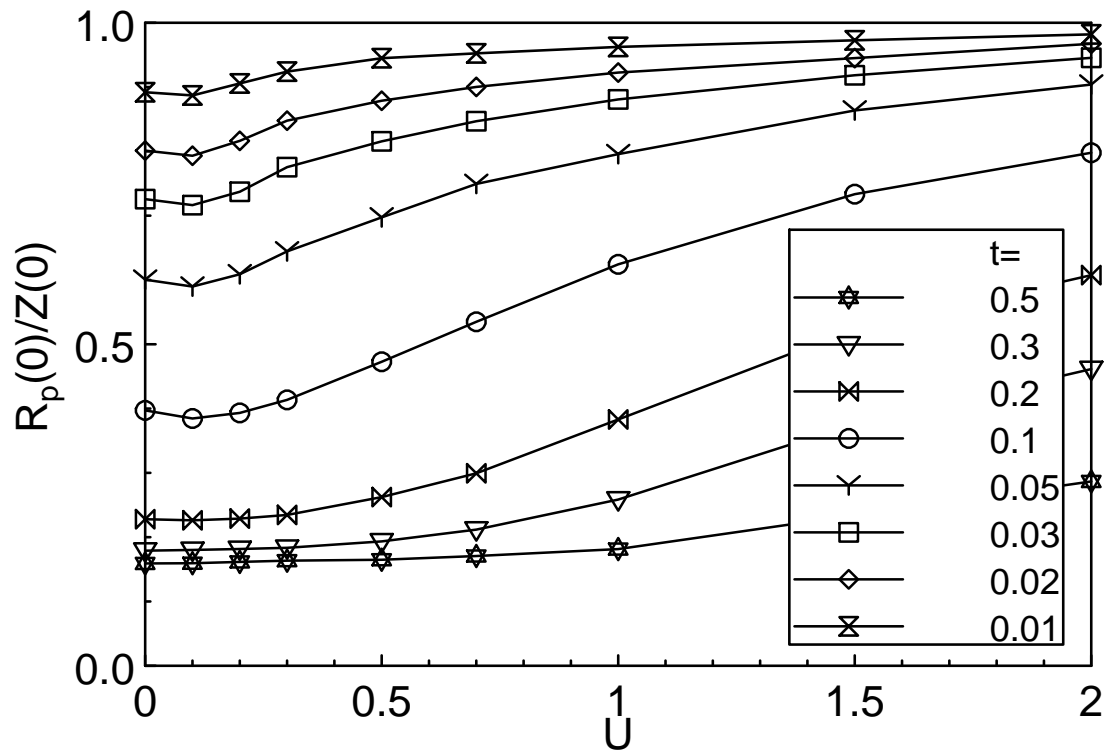}}
  \caption{Normalized single-particle return probability at the Fermi energy,
  $R_\sigma(0)/Z_\sigma(0)$; parameters as in Fig. \protect\ref{Fig:return}.}
  \label{Fig:ret_norm}
\end{figure}

In order to extract the localization  information from
the return probability, we normalize $R_\sigma$ by $Z_\sigma$.
The data in Fig. \ref{Fig:ret_norm} show that, in general, interactions
tend to localize the {\em single-particle} excitations. The delocalization
at low interaction strength and small kinetic energy is only a tiny effect.
This is in agreement with earlier results (Epperlein {\it et al.} 1997)
based on
the Hartree-Fock approximation. The reason for the strong single-particle
localization is the Coulomb gap in the {\em single-particle} density of
states which effectively reduces the overlap between excitations close to the
Fermi level.

\begin{figure}
  \epsfxsize\figuresize
  \centerline{\epsffile{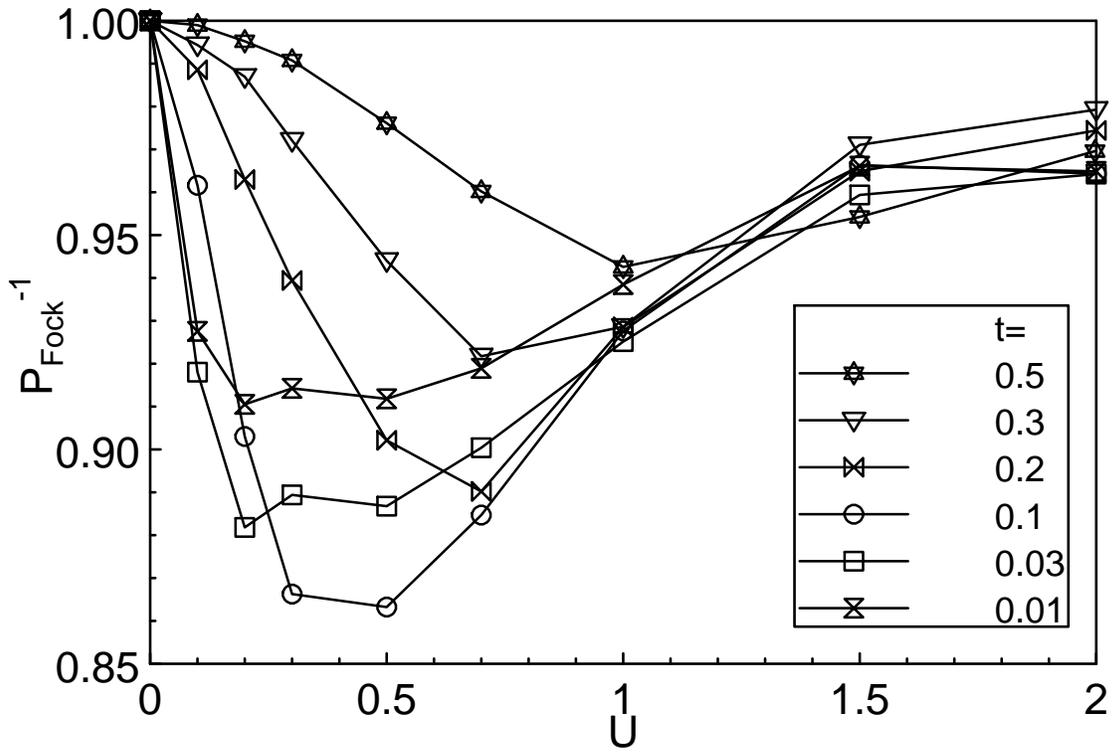}}
  \caption{Inverse Fock space participation number $P_{\rm Fock}^{-1}$
      as a function of interaction $U$; parameters as in in Fig. \protect\ref{Fig:return}.}
  \label{Fig:fockpn}
\end{figure}

For comparison we also calculate
the inverse participation number of the many-particle ground state with
respect to the Hartree-Fock Slater determinants. This quantity
measures, to what extent the interactions introduce non-trivial
correlations (beyond the Hartree-Fock level) into the system.
Conceptually, it is therefore similar to the survival probability (\ref{eq:survive}).
Our results are shown in Fig. \ref{Fig:fockpn}.
For $U=0$ the Fock space participation number is 1 by definition
because the Hartree-Fock approximation is exact for non-interacting electrons.
Increasing
the interaction mixes the single-particle states, and the inverse Fock space
participation number is reduced. For large interaction the eigenstates
become again more localized in Fock space. Finally, for $U \to \infty$
the Hartree-Fock approximation becomes exact, and the inverse Fock space
participation number approaches 1 again. The delocalization in Fock space
roughly occurs in the same parameter region where the conductance is enhanced
by interactions.

\section{Summary and conclusions}
\label{sec:SUM}

In this paper we have studied the transport and localization properties
of interacting spinless fermions and electrons in a random potential.
The numerical calculations have been carried out by means of the
Hartree-Fock based diagonalization method.

Our results for the influence of interactions on the conductance, viz. a
localizing tendency in the diffusive regime and a delocalizing tendency
deep in the insulating regime, are in qualitative agreement with a number
of other studies in one and two dimensions.
We do not see any indications for a true metal-insulator transition in
the systems considered. Of course, new phenomena could develop at
significantly larger length scales. However, the energy scale for
the 2D MIT observed experimentally in the Si-MOSFETs is the Fermi energy
rather than some additional low-energy scale. Therefore one would expect
to see signatures of this transition already at small length scales.

We have paid particular attention to the spin degrees of freedom.
To this end we have compared the conductance values of spinless fermions
and electrons, both systems being at half filling. We have found that
the interaction dependence of the conductance is much stronger for
electrons than for spinless fermions. However, this stronger dependence can
simply be accounted for by rescaling the Coulomb interaction strength
by 2. Therefore, spin phenomena do not seem to play an essential role
here, in contrast to the experiments in Si-MOSFETs where an
in-plane magnetic field (which does not couple to the orbital motion of
the electrons) strongly suppresses the conducting phase
(Popovich {\it et al.} 1997, Simonian {\it et al.} 1997).

In addition to the conductance we have studied the localization
of single-particle excitations, which we have characterized by the normalized
return probability, i.e. the return probability divided by the square
of the quasiparticle weight. Single-particle excitations generally tend
to become more localized under the influence of the interactions.
The reason for the strong single-particle
localization is the Coulomb gap in the {\em single-particle} density of
states which effectively reduces the overlap between excitations close to the
Fermi level. In contrast, the particle-hole excitations responsible for the
conductance do not have a Coulomb gap, and thus they show a tendency towards
delocalization at low interaction strength and small kinetic energy.

\section*{Acknowledgements}

This paper is dedicated to Prof.\ Michael Pollak on the occasion of his
75th birthday. Michael Pollak has greatly enriched the field of disorder
and interactions in the insulating phase. His contributions have
spanned more than 30 years, starting with the pioneering discovery of
the classical Coulomb gap in 1970 and continuing till today's work on
quantum localization and interactions.

This work was supported in part by the German Research Foundation
under Grant No. SFB393/C2.
\vspace{5mm}

\frenchspacing
\parindent0pt
{\large\bf References}\\

Altshuler, B.L. and Aronov, A.G., 1985, in {\it Electron-electron
      interactions in disordered systems}, edited by A.L. Efros and M. Pollak
      (Amsterdam: North Holland)\\

Altshuler, B.L. and Maslov, D.L., 1999, Phys. Rev. Lett. {\bf 82} 145\\

Belitz, D. and Kirkpatrick, T.R., 1994, Rev. Mod. Phys. {\bf 66}, 261\\

Belitz, D. and Kirkpatrick, T.R., 1998, Phys. Rev. B {\bf 58}, 8214\\

Belitz, D. and Kirkpatrick, T.R., 1999, Ann. Phys. (Leipzig) {\bf 8}, 765\\

Benenti, G., Waintal, X., and Pichard, J.-L., 1999,
      Phys. Rev. Lett. {\bf 83}, 1826\\

Berkovits, R. and Avishai, Y., 1996, Phys. Rev. Lett. {\bf 76}, 291\\

Berkovits, R., Kantelhardt, J.W., Avishai, Y., Havlin, S., and Bunde, A.,
     2001,  Phys. Rev. B {\bf 63}, 085102\\

Castellani, C., DiCastro, C., and Lee, P.A., 1998, Phys. Rev. B {\bf 57},
      R9381\\

Chakravarty, S., Kivelson, S., Nayak, C., and Voelker, K., 1999,
      Phil. Mag. B {\bf 79}, 859\\

Dagotto, E., 1994, Rev. Mod. Phys. {\bf 66}, 763\\

Das Sarma, S. and Hwang, E.H., 1999, Phys. Rev. Lett. {\bf 83}, 164 \\

Datta, S., 1997, {\it Electronic transport in
    mesoscopic systems} (Cambridge:  Cambridge University Press)\\

Denteneer, P.J.H., Scalettar, R.T., and Trivedi, N., 1999, Phys. Rev. Lett.
      {\bf 83}, 4610\\

Economou, E.N. and Cohen, M.H., 1970,  Phys. Rev. Lett. {\bf 25},
    1445\\

Efros, A.L. and Pikus, F.G., 1995,  Solid State Commun.
   {\bf 96}, 183\\

Efros, A.L. and Pollak M., 1985, {\it Electron-electron interactions
       in disordered systems} (Amsterdam: North-Holland)\\

Efros, A.L. and Shklovskii, B.I., 1975, J. Phys. C {\bf 8}, L49\\

Epperlein, F., Schreiber, M., and Vojta, T., 1997, Phys. Rev. B {\bf 56},
    5890\\

Epperlein, F., Schreiber, M., and Vojta, T., 1998, phys. stat. sol. (b) {\bf
    205}, 233\\

Finkelstein, A.M., 1983, Zh. Eksp. Teor. Fiz. {\bf 84}, 168
      [Sov. Phys. JETP {\bf 57}, 97 (1983)]\\

Fulde, P., 1995, {\it Electron correlations in molecules and solids}
       (Berlin: Springer)\\

Gold, A. and Dolgopolov, V.T., 1986, Phys. Rev. B {\bf 33}, 1076\\

Greenwood, D.A., 1958, Proc. Phys. Soc. {\bf 71}, 585\\

Klapwijk, T.M.  and Das Sarma, S., 1999, Solid State Commun. {\bf 110}, 581\\

Kravchenko, S.V., Kravchenko, G.V., Furneaux, J.E., Pudalov, V.M.,
      and D'Iorio, M., 1994, Phys. Rev. B {\bf 50}, 8039\\

Kravchenko, S.V., Mason, W.E., Bowker, G.E., Furneaux, J.E.,
      Pudalov, V.M., and D'Iorio, M., 1995, Phys. Rev. B {\bf 51}, 7038\\

Kubo, K., 1957, J. Phys. Soc. Jpn. {\bf 12}, 576\\

Lee, P.A. and Ramakrishnan, T.V., 1985, Rev. Mod. Phys. {\bf 57}, 287\\

M\"obius, A., Neklioudov, A., Diaz-Sanchez, A., Hoffmann, K.H., Fachat,
       A., and Schreiber, M., 1997, Phys. Rev. Lett. {\bf 79}, 4297\\

M\"obius, A., Richter, M., and Drittler, B., 1992, Phys. Rev. B {\bf 45},
     11568\\

Mott, N.F., 1968, J. Non-Cryst. Sol. {\bf 1}, 1\\

Phillips, P., Wan, Y., Martin, I., Knysh, S., and Dalidovich, D., 1998,
      Nature {\bf 395}, 253\\

Pollak, M., 1970,  Discuss. Faraday Soc. {\bf 50}, 13\\

Pollak, M., 1992, Phil. Mag. B {\bf 65}, 855\\

Popovich, D., Fowler, A.B., and Washburn, S., 1997, Phys. Rev. Lett. {\bf 79},
    1543\\

R\"omer, R.A., Schreiber, M., and Vojta, T., 2001, Physica E {\bf 9}, 397\\

Sarvestani, M., Schreiber, M., and Vojta, T., 1995,
     Phys. Rev. B {\bf 52}, R3820\\

Schmitteckert, P., Jalabert, R.A., Weinmann, D., and  Pichard, J.-L., 1998,
      Phys. Rev. Lett. {\bf 81}, 2308\\

Schreiber, M., Epperlein, F., and Vojta, T., 1999, Physica A
    {\bf 266}, 443\\

Shepelyansky, D.L., 1994, Phys. Rev. Lett. {\bf 73}, 2607\\

Si, Q. and Varma, C.M., 1998, Phys. Rev. Lett. {\bf 81}, 4951\\

Simonian, D., Kravchenko, S.V., and Sarachik, M.P., 1997,
    Phys. Rev. Lett. {\bf 79}, 2304\\

Talamantes, J., Pollak, M., and Elam, L.,  1996, Europhys. Lett.
   {\bf 35}, 511\\

Vojta, T. and Epperlein, F., 1998, Ann. Phys. (Leipzig) {\bf 7}, 493\\

Vojta, T., Epperlein, F., and Schreiber, M., 1998a, phys. stat. sol. (b)
    {\bf 205}, 53\\

Vojta, T., Epperlein, F., and Schreiber, M., 1998b, Phys. Rev. Lett.
    {\bf 81}, 4212\\

Vojta, T., Epperlein, F.,  and Schreiber, M., 1999,
    Comp. Phys. Commun. {\bf 121--122}, 489\\

von der Linden, W., 1992, Phys. Rep. {\bf 220}, 53\\

White, S.R., 1998, Phys. Rep. {\bf 301}, 187\\

Zhang, F.C. and Rice, T.M., 1997, cond-mat/9708050\\

\newpage
\newpage

\newpage

\newpage

\newpage

\newpage

\end{document}